# Enhancing the blocking temperature of perpendicular-exchange biased $Cr_2O_3$ thin films using spacer and buffer layers

**Naoki Shimomura[1], Satya Prakash Pati[1], Tomohiro Nozaki[1], Tatsuo Shibata[2], and Masashi Sahashi[1]**

[1]Department of Electronic Engineering, Graduate School of Electronic Engineering, Tohoku University, Sendai 980-8579, Japan

[2]TDK Corporation, Ichikawa 272-8558, Japan

E-mail: nozaki@ecei.tohoku.ac.jp

In this study, we investigated the effect of spacer and buffer layers on the blocking temperature $T_B$ of the perpendicular exchange bias of thin $Cr_2O_3$ films, and revealed a high $T_B$ of 260 K for 20-nm-thick $Cr_2O_3$ thin films. By inserting a Ru spacer layer between the $Cr_2O_3$ and Co films and changing the spacer thickness, we controlled the magnitude of the exchange bias and $T_B$. By comparing the $T_B$ values of the 20-nm-thick $Cr_2O_3$ films on Pt and $\alpha$-$Fe_2O_3$ buffers, we investigated the lattice strain effect on the $T_B$. We show that higher $T_B$ value can be obtained using an $\alpha$-$Fe_2O_3$ buffer, which is likely because of the lattice-strain-induced increase of $Cr_2O_3$ magnetic anisotropy.

## 1. Introduction

Electric control of magnetization using the magnetoelectric (ME) effect has received considerable attention as a promising candidate for next generation low energy consumption magnetic recording devices[1]. The linear ME effect was theoretically predicted for $Cr_2O_3$ in 1960[2] and experimentally confirmed in 1961[3]. Few techniques at that time had utilized to deal antiferromagnets, so the practical potential of the discovery was not immediately realized. Since 2005, the ME effect in $Cr_2O_3$ has captured renewed attention because the electrical switching of perpendicular exchange bias has been demonstrated in bulk $Cr_2O_3$ single crystal/ferromagnet exchange coupling systems[4,5]. Here, antiferromagnetic domain reversal by applying both magnetic and electric fields[6] has been used to switch the perpendicular exchange bias of a $Cr_2O_3$/ferromagnet. This perpendicular exchange bias switching has recently been demonstrated in several-hundred-nanometer-thick $Cr_2O_3$ films deposited by the sputtering method, which yields more realistic device applications[7-11]. One major issue for device application is simultaneous realization of both further reduction of the $Cr_2O_3$ film thickness and increase in the operating temperature. Further reduction of the $Cr_2O_3$ film thickness is necessary to decrease the applied voltage and the aspect ratio of the recording bit. However, for thin $Cr_2O_3$ films, there is a rapid reduction in the blocking temperature $T_B$ of the perpendicular exchange bias with decreasing $Cr_2O_3$ thickness[12]. Here, $T_B$ represents the temperature at which the perpendicular exchange bias disappears and it is the upper limit of the operating temperature. Thus, in addition to the enhancement of the Néel temperature $T_N$, for which several attempts have been reported recently[13-16], to achieve high $T_B$ is also an important issue. Perpendicular exchange bias has been reported for $Cr_2O_3$ film thicknesses $\geq$ 30 nm[12]. However, the $T_B$ is as low as 80 K for 30-nm-thick $Cr_2O_3$ film when the unidirectional magnetic anisotropy energy $J_K$ (=$H_{ex}M_st_{FM}$) is 0.26 mJ/m$^2$. Here, $H_{ex}$, $M_s$, and $t_{FM}$ represent the exchange bias field, saturation magnetization and thickness of the ferromagnetic layer, respectively. If the $Cr_2O_3$ thickness is further decreased, the perpendicular exchange bias is not observed even at very low temperatures. For $Cr_2O_3$, the lower $T_B$ than $T_N$ (~307 K) can be qualitatively understood by using the Meiklejohn–Bean free-energy model (MB model)[17,18]. In the MB model, the $T_B$ is explained by the competition between the interface exchange coupling energy $J_{ex}$ and the product of the magnetic anisotropy energy $K_{AF}$ and thickness $t_{AF}$ of the antiferromagnet ($K_{AF}t_{AF}$). When $K_{AF}t_{AF} > J_{ex}$, the exchange bias is observed (unidirectional anisotropy). When $K_{AF}t_{AF} < J_{ex}$, the exchange bias disappears and only the enhancement of the coercivity $H_c$ is observed (uniaxial anisotropy). Such appearance/disappearance of the exchange bias has also been observed for antiferromagnetic Mn–Ir alloy[19]. For the Mn–Ir alloy, the exchange bias at room temperature disappears and the $H_c$ increases when the Mn–Ir thickness decrease below ~5 nm. That is, for a 5-nm-thick Mn–Ir film, the $T_B$ is approximately equal to the room temperature. For $Cr_2O_3$, because of its small $K_{AF}$ (~$2.0 \times 10^4$ $J/m^3$ at low temperature[20]) and large $J_{ex}$, the $T_B$ becomes much lower than the $T_N$ even for a 500-nm-thick $Cr_2O_3$ film. The $T_B$ of the $Cr_2O_3$/Co exchange coupling system is easily controlled by changing the $t_{AF}$ or $J_{ex}$[18], which indicates that the $T_B$ of the $Cr_2O_3$/Co system can be well described by the MB model. According to the MB model, to realize a high $T_B$ in thin $Cr_2O_3$ layers, either the $J_{ex}$ needs to be decreased or the $K_{AF}$ needs to be increased. Decreasing the $J_{ex}$ can be achieved by inserting a thin metallic spacer between the $Cr_2O_3$ film and ferromagnet. Increasing $T_B$ by inserting a Pt spacer layer has been reported, where the reduction in the $J_{ex}$ was confirmed as a reduction in the $H_{ex}$[18,21,22]. A higher $K_{AF}$ can be achieved by doping and inducing lattice strain. The $K_{AF}$ consists of magnetic dipole anisotropy $K_{MD}$ and magnetocrystalline anisotropy (the fine structure anisotropy) $K_{FS}$. Increasing the $K_{AF}$ by Al-doping has been reported, which is mainly mediated by increasing the $K_{FS}$[23]. Artman et al. calculated the change in the $K_{MD}$ induced by the lattice parameter variation and found that the $K_{MD}$ increases with increasing $c$, decreasing $a$, or increasing ionic position $w$[20]. The $K_{FS}$ may also be affected by the lattice strain, but this effect has not been quantified. Because lattice strain can be easily induced for thin films by adjusting the lattice parameter of the buffer layer, a thin $Cr_2O_3$/Co system is favorable for investigating the effect of the lattice strain on the $K_{AF}$. In this study, we attempted to achieve $T_B \approx T_N$ for a 20-nm-thick $Cr_2O_3$ thin film through controlling the $J_{ex}$ by inserting a metallic spacer layer and enhancing the $K_{AF}$ by inducing lattice strain.

## 2. Experimental details

All the samples were fabricated by RF-DC magnetron sputtering with a base pressure below $5 \times 10^{-7}$ Pa. The sample structures were c-$Al_2O_3$ substrate/Pt 25 or α-$Fe_2O_3$ 20/$Cr_2O_3$ 20/Ru $t_{Ru}$/Co 1/Pt 5 (nm). In this study, we varied the Ru spacer layer thickness $t_{Ru}$ and buffer layer material (Pt or α-$Fe_2O_3$). The oxygen reactive sputtering technique was used for deposition of $Cr_2O_3$ (α-$Fe_2O_3$) by sputtering the Cr (Fe) metal target in an Ar/$O_2$ mixed gas atmosphere. The substrate temperature during deposition was 873 K for the Pt buffer layer, 773 K for the $Cr_2O_3$ and α-$Fe_2O_3$ oxide layers, and 406 K for the Ru spacer, Co ferromagnetic, and Pt capping layers. The magnetic properties were measured by superconducting quantum interference device (SQUID) magnetometry after cooling the sample in the presence of an applied magnetic field (+10 kOe) from 340 K, which is sufficiently above the $T_N$ of $Cr_2O_3$. During the measurements, the magnetic field was applied normal to the film surface. Structural characterization was performed by X-ray diffraction (XRD) and cross-sectional transmission electron microscopy (TEM) measurements. Nanobeam electron diffraction was used for refining the lattice parameters of the thin $Cr_2O_3$ layer (20 nm thick)[24].

**3. Ru spacer layer effect**

First, we investigated the effect of the Ru spacer layer on the $H_{ex}$ using Pt-buffered samples. We measured the dependence of the $H_{ex}$ on the Ru spacer layer thickness for the Ru spacer samples ($Al_2O_3$/Pt buffer 25/$Cr_2O_3$ 250/Ru spacer $t_{Ru}$/Co 1/Pt 5), and compared the results with those of the Pt spacer samples ($Al_2O_3$/Pt buffer 25/$Cr_2O_3$ 250/Pt spacer $t_{Pt}$/Co 1/Pt 5). In thin $Cr_2O_3$ films, because of the small $K_{AF}t_{AF}$, only the $H_c$ enhancement will occur, making these films unsuitable for the $H_{ex}$ comparison. Thus, we used relatively thick $Cr_2O_3$ films (250-nm-thick). Figure 1(a) shows the dependence of the $J_K$ on the $t_{Ru}$ at 50 K. For comparison, the results for the Pt spacer samples[22] are also shown in figure 1(a). For the $J_K$ calculations, the spacer layer thickness dependence of $M_st_{FM}$ was measured for both the Pt and Ru spacer samples. The $M_st_{FM}$ values of 1.70(17) × $10^{-4}$ emu/cm$^2$ for the no-spacer and Pt spacer samples, and 1.34(9) × $10^{-4}$ emu/cm$^2$ for the Ru spacer samples are obtained and its spacer thickness dependence are within the error. Due to the additional moment mainly comes from spin polarization of the 5 nm Pt cap layer, the no-spacer and Pt-spacer samples exhibit higher $M_st_{FM}$ than that of 1 nm Co (≈1.4 × $10^{-4}$ emu/cm$^2$). The Ru spacer samples exhibit lower $M_st_{FM}$ values than the no-spacer and Pt spacer samples. It seems the spin polarization of the 5 nm Pt cap layer is suppressed by Ru spacer, possibly due to the change in the crystal structure of Co and Pt or change in the magnetic anisotropy of Co by using a Ru spacer layer. Figure 1(b) shows the $M$–$H$ curves of the Ru and Pt spacer samples at 50 K. Compared to the Pt spacer samples, the Ru spacer samples exhibit lower $H_c$ and lower squareness. There are two main effects of the spacer layer: the effect on the magnitude of exchange coupling and the effect on the magnetic properties of Co, especially the magnetic anisotropy of Co. Regarding the magnitude of exchange coupling, the $H_{ex}$ generally decreases with increasing spacer layer thickness. The spacer layer lengthen the $Cr_2O_3$-Co distance and weaken the $J_{ex}$. Thus the magnitude of exchange coupling ($H_{ex}$ and $J_K$) decrease. This is true for both the Ru and Pt spacers, although there is a slight difference in the thickness dependence. For the Pt spacer samples, there is a rapid decrease in the $J_K$ with increasing $t_{Pt}$ for $t_{Pt}$ > 0.7 nm, while for the Ru spacer samples, the $J_K$ linearly decreases with increasing $t_{Ru}$. From these results, we assume the island growth of Pt for the Pt spacer samples when the Pt spacer layer is thin ($t_{Pt}$ < 0.7 nm), which leads to a smaller change in the $J_{K,}$ reflecting the small change in the $Cr_2O_3$-Co distance. In contrast, the layer growth is expected for Ru of the Ru spacer sample, which leads to a gradual change in $Cr_2O_3$-Co distance and $J_{K,}$ reflecting the linear change in the $Cr_2O_3$-Co distance. It is noted that the thickness dependence is affected not only by the spacer material but also by the spacer layer deposition conditions (e.g., comparing the results of refs. 14, 18 and 25). These results show that the $J_K$ of $Cr_2O_3$/Co films can be controlled more easily by using Ru spacers than Pt spacers. On the other hand, the changes in the magnetic properties of Co largely depend on the spacer layer material (Pt or Ru), and they are independent of the spacer thickness ($t_{Pt}$ or $t_{Ru}$). A larger $H_c$ value and higher squareness of the Pt spacer sample indicate interface perpendicular magnetic anisotropy at the Pt/Co interface. Although the magnitudes of the magnetic anisotropy of Co differ between the Pt and Ru spacer samples, the affection on the spacer thickness dependence of the $J_K$ seems limited. For the Pt spacer samples, a rapid change in $J_K$ occur when $t_{Pt}$ > 0.7 nm, while the magnetic anisotropy of Co is almost unchanged in this region. For the Ru spacer samples, no abrupt change was observed for $J_K$ values, while the magnetic anisotropy dramatically change between $t_{Ru}$ = 0.0 to $t_{Ru}$ = 0.25 nm.

Here, we demonstrate the enhancement of the $T_B$ for a 20-nm-thick $Cr_2O_3$ film by inserting a Ru spacer layer. Figure 2 shows the dependence of the $H_{ex}$ and $H_c$ on temperature for c-$Al_2O_3$/Pt 25/$Cr_2O_3$ 20/Ru $t_{Ru}$/Co 1/Pt 5 (nm). In the sample without the Ru spacer, only the $H_c$ enhancement is observed and no $H_{ex}$ is observed from 10 to 300 K (black open triangles), owing to large $J_{ex}$ and small $K_{AF}t_{AF}$. In contrast, in the samples with $t_{Ru}$ > 1 nm (red solid squares and blue solid circles), where the $H_{ex}$ < 1 kOe ($J_K$ < 0.134 mJ/m$^2$), the $H_{ex}$ is observed at low temperatures. For the sample with $t_{Ru}$ = 1.25 nm, a relatively high $T_B$ of 150 K is attained.

**4. Buffer layer effect**

We investigated the effect of lattice strain induced by changing the buffer layer material on the $K_{AF}$ of $Cr_2O_3$. For a $Cr_2O_3$ buffer layer, good lattice matching and oxidative resistance at 773 K (deposition temperature of $Cr_2O_3$) are required. In this study, we chose α-$Fe_2O_3$ and Pt for the buffer layer materials. Both $Cr_2O_3$ and α-$Fe_2O_3$ possess the corundum structure, so an epitaxial growth of α-$Fe_2O_3$/$Cr_2O_3$ is expected. Pt is a face-centered cubic structure, but it preferentially orients along the [111] direction. Thus, the lattice matching is relatively good. In addition, both α-$Fe_2O_3$ and Pt show good oxidative resistances. Figure 3 shows the schematics of the (0001) planes of $Cr_2O_3$ and $\alpha$-$Fe_2O_3$, and the (111) plane of Pt. Table 1 summarizes the lattice mismatch between $Cr_2O_3$ and the buffer layers. Because $\alpha$-$Fe_2O_3$ (Pt) has larger (smaller) lattice parameters than $Cr_2O_3$, we expected to obtain *a*-axis expanded (compressed) $Cr_2O_3$ using an (a) $\alpha$-$Fe_2O_3$ (Pt) buffer. We investigated the change in the $K_{AF}$ of these samples by measuring the $T_B$, and compared the results with the theoretical predictions.

### *4.1. Structural characterization*

The lattice parameters and morphology can be affected by changing the buffer layer. Thus, we first characterized the structural properties. Figure 4 shows the XRD patterns for $2\theta/\omega$ (out of plane, figure 4(a)) and $2\theta_\chi/\phi$ (in plane, figure 4(b)) scans of $\alpha$-$Fe_2O_3$-buffered and Pt-buffered samples. The $2\theta_\chi/\phi$ scan was carried out for the $Al_2O_3$ $(10\bar{1}0)$ plane. For the $\alpha$-$Fe_2O_3$-buffered sample, a good epitaxial growth was observed. The epitaxial relations are $Al_2O_3$ (0001) $[10\bar{1}0]$/$\alpha$-$Fe_2O_3$ (0001) $[10\bar{1}0]$/$Cr_2O_3$ (0001) $[10\bar{1}0]$. For the Pt-buffered sample, although good (111) oriented Pt and (0001) oriented $Cr_2O_3$ were obtained, two types of domains appeared in the in-plane orientations. The epitaxial relations are $Al_2O_3$ (0001) $[10\bar{1}0]$/Pt (111) $[1\bar{1}0]$ or $[1\bar{1}2]$/$Cr_2O_3$ (0001) $[10\bar{1}0]$ or $[11\bar{2}0]$. Figure 5 shows the cross-sectional TEM images of the $\alpha$-$Fe_2O_3$-buffered (figure 5(a)) and Pt-buffered (figure 5(b)) samples. The TEM image confirmed the epitaxial growth of the $\alpha$-$Fe_2O_3$-buffered sample and existence of the two types of domains in the Pt-buffered sample (domain sizes of several tens of nanometers). The different domain sizes can also affect the $T_B$. However, based on the MB model, the domain sizes perpendicular to the film surface are more important than the domain sizes in the film plane, because we discuss the relation between $J_K$ and $K_{AF}t_{AF}$. In this paper, we neglect the grain size effect, and additional investigations are underway. We then checked the lattice parameter *a* of $Cr_2O_3$ on $\alpha$-$Fe_2O_3$ and Pt. We have previously reported the *a* value of $\alpha$-$Fe_2O_3$- and Pt-buffered 20-nm-thick $Cr_2O_3$ films from the XRD patterns[26). However, the *a* value of the $\alpha$-$Fe_2O_3$-buffered sample could not be accurately determined owing to the overlap between the $Cr_2O_3$ $(30\bar{3}0)$ and $\alpha$-$Fe_2O_3$ $(30\bar{3}0)$ Bragg peaks. In the present study, we performed nanobeam electron diffraction measurements[24) of the $\alpha$-$Fe_2O_3$-buffered sample and more precisely determined the *a* value of the $Cr_2O_3$ film. The *a* value of $Cr_2O_3$ estimated from the nanobeam electron diffraction was the same as that of the $\alpha$-$Fe_2O_3$ buffer estimated from in-plane XRD. Thus, we confirmed that the *a* value of $Cr_2O_3$ is locked by the $\alpha$-$Fe_2O_3$ buffer layer. From this result, we could identify that both the $\alpha$-$Fe_2O_3$ $(30\bar{3}0)$ and $Cr_2O_3$ $(30\bar{3}0)$ peaks are at 64.1° in figure 4(b). The refined lattice parameter values are listed in table 1. Both the present data and the data in[26] show that the *a* values for α-$Fe_2O_3$-buffered $Cr_2O_3$ films are larger than that of the Pt-buffered $Cr_2O_3$ film, and both the samples have larger *a* values than bulk $Cr_2O_3$ ($a$ = 4.95 Å). The unexpected *a* value of the Pt-buffered $Cr_2O_3$ could have resulted from the lattice relaxation due to misfit dislocations or dislocations at grain boundaries at Pt/$Cr_2O_3$ interface owing to a relatively bad lattice matching, while it is difficult to identify the dislocations from TEM images (Fig. 5 (b)).

*4.2. Blocking temperature and magnetic anisotropy*

Next, using 1.25-nm-thick Ru spacer samples, for which finite $H_{ex}$ and $T_B$ were obtained, we clarified the buffer layer effect on the $T_B$ using Pt- and α-$Fe_2O_3$-buffered sample with 20-nm-thick $Cr_2O_3$. Figure 6 shows the temperature dependence of the $J_K$ for Pt- and α-$Fe_2O_3$-buffered 20-nm-thick $Cr_2O_3$ (c-$Al_2O_3$/Pt 25 or α-$Fe_2O_3$ 20/$Cr_2O_3$ 20/Ru 1.25/Co 1/Pt 5 (nm)). As shown in figure 2, $T_B \approx 150$ K for the Pt-buffered sample. A much higher $T_B \approx 260$ K was obtained for the α-$Fe_2O_3$-buffered samples. Since the $T_N$ of the α-$Fe_2O_3$-buffered 20-nm-thick $Cr_2O_3$ film was lowered to 269 K due to the lattice strain[26], we managed to obtain $T_B \approx T_N$ for a 20-nm-thick $Cr_2O_3$ film. Because the $J_K$ values of these samples are almost equal, or slightly larger for the α-$Fe_2O_3$-buffered samples, the $K_{AF}t_{AF}$ of the α-$Fe_2O_3$-buffered samples must be much higher than that of the Pt-buffered samples. In other words, the $K_{AF}$ of the $Cr_2O_3$ films increased by using the α-$Fe_2O_3$ buffer layer. If we assume the Mauri's domain wall model[27], the higher $K_{AF}$ links to the higher $H_{ex}$($J_K$). The slightly larger $J_K$ of α-$Fe_2O_3$-buffered samples can comes from the higher $K_{AF}$, while more works are required for the confirmations. Based on the Mauri's model, the $T_N$ change also affect to the magnitude of the $H_{ex}$. However the affection of $H_{ex}$ by the less than 10% difference in $T_N$ between α-$Fe_2O_3$-buffered sample ($T_N \sim 269$ K) and Pt-buffered sample ($T_N \sim 294$ K) should be negligible small. In addition, we investigated the relations between the $J_K$ and $T_B$ for the Pt- and α-$Fe_2O_3$-buffered samples with different Ru spacer thicknesses. Figure 7 shows the $T_B$ values of the Pt- and α-$Fe_2O_3$-buffered samples plotted against the $J_K$ at 50 K. Irrespective of the Ru spacer thickness, higher $T_B$ values were obtained for the α-$Fe_2O_3$-buffered samples, while the $T_B$ decreased with increasing $J_K$ for both samples. These results clearly demonstrate that the $K_{AF}$ of the $Cr_2O_3$ layer is higher when using an α-$Fe_2O_3$ buffer layer than a Pt buffer layer.

Moreover, we investigated the $Cr_2O_3$ layer thickness dependence of the $J_K$ and $T_B$ for the α-$Fe_2O_3$-buffered sample. Figure 8 shows the temperature dependence of the $J_K$ for $Al_2O_3$/α-$Fe_2O_3$ 20/$Cr_2O_3$ $t_{Cr2O3}$/Ru 1.25/Co 1/Pt 5 (nm) with various $Cr_2O_3$ thicknesses (3 nm ≤ $t_{Cr2O3}$ ≤ 20 nm). We observed an exchange bias for a 5-nm-thick $Cr_2O_3$ sample ($T_B$ ≈ 10 K), although thinner $Cr_2O_3$ samples exhibited no apparent exchange bias. As expected from the MB model, the $T_B$ decreases with decreasing $Cr_2O_3$ thickness, while the magnitude of the $J_K$ is almost unchanged in the 5 nm ≤ $t_{Cr2O3}$ ≤ 20 nm range. These results indicate that the MB model is qualitatively applicable for this system even for thin $Cr_2O_3$ regions ($t_{Cr2O3}$ ≤ 20 nm).

Because α-$Fe_2O_3$ is an antiferromagnet, in addition to the lattice strain effect, an enhancement of $K_{AF}$ due to the interlayer coupling between the antiferromagnetic α-$Fe_2O_3$ and $Cr_2O_3$ can be considered, as reported in the NiO/CoO system[28]. However, the observed dependence of the $T_B$ on the $Cr_2O_3$ thickness does not support this assumption. If the interlayer coupling effect is dominant, the $T_B$ will not decrease with decreasing $Cr_2O_3$ thickness because the α-$Fe_2O_3$ thickness is constant. However, our data suggests that the $T_B$ decreases with decreasing $Cr_2O_3$ thickness. Thus, the interlayer coupling effect is negligibly small in this case, probably owing to the small $K_{AF}$ of α-$Fe_2O_3$ (~2 × $10^4$ J/$m^3$ at low temperature[20]).

Using the MB model, we estimated the change in the $K_{AF}$ for 20-nm-thick $Cr_2O_3$ films due to the strain induced by the buffer layer. According to the MB model, at the critical point where the exchange bias abruptly disappears, the relationship $K_{AF} = J_{ex}/t_{AF}$ holds true. If we assume that the $J_{ex}$ is almost the same as the $J_K$, we can estimate the $K_{AF}$ as $K_{AF} = J_K/t_{AF}$. In fact, the $K_{AF}$ values of FeMn/NiFe[29] and IrMn/NiFe[19] at room temperature have been estimated from $K_{AF} = J_K/t_{AF}^{cr}$ by determining the critical antiferromagnet thickness $t_{AF}^{cr}$. In this study, we estimated $K_{AF} = J_K^{cr}/t_{AF}$. We determined the critical unidirectional magnetic anisotropy energy $J_K^{cr}$ at 100 K by changing the Ru spacer layer thickness for $Al_2O_3$/Pt 25 or α-$Fe_2O_3$ 20/$Cr_2O_3$ 20/Ru $t_{Ru}$/Co 1/Pt 5 (nm) structure sample. The critical Ru spacer layer thickness was 0.75 nm (1.25 nm) for the α-$Fe_2O_3$- (Pt-) buffered sample, and $J_K^{cr}$ = 0.37 mJ/$m^2$ (0.09 mJ/$m^2$) was obtained, as shown in Fig. 9. The estimated $K_{AF}$ values at 100 K were 1.9 × $10^4$ J/$m^3$ for the α-$Fe_2O_3$-buffered sample and 4.5 × $10^3$ J/$m^3$ for the Pt-buffered sample. Note that the $T_B$ of these samples are not exactly 100 K, but between 100 K and 150 K. Due to the rough determination of $T_B$, the $J_K^{cr}$ and the estimated $K_{AF}$ values are slightly underestimated. Because the calculation of the $K_{AF}$ using the MB model includes many assumptions, we could not obtain an exact absolute value of the $K_{AF}$ from these calculations. However, it can be used to compare the two samples with similar structures. In this study, only the buffer layers were different. The other film properties such as the $Cr_2O_3$ layer thickness, spacer layer, and Co layer thickness were maintained. Because the precise characterization of the $K_{AF}$ of antiferromagnetic thin films is considerably difficult, we believe that this is a good method to approximately estimate the $K_{AF}$. Based on these concepts, it was found that the $K_{AF}$ of the α-$Fe_2O_3$-buffered sample is nearly four times that of the Pt-buffered sample.

Here, we compare the experimental results and theoretical predictions assuming that the variation of the $K_{AF}$ mainly originates from the lattice strain. The experimental $a$ values of $Cr_2O_3$ and $T_B$ values are summarized in table 2. In our results, the $T_B$ increases with increasing $a$, indicating increasing $K_{AF}$ of $Cr_2O_3$. Because Artman et al. predicted that the $K_{MD}$ should decrease with increasing $a$[20)], our experimental results cannot be explained based only on the lattice-strain-induced $K_{MD}$ changes. We believe that other effects, such as the lattice strain effect on the $K_{FS}$ or the effect of $w$ on the $K_{AF}$, will mainly contribute to the change in the $K_{AF}$. In fact, for Al-doped $Cr_2O_3$, the change in the $K_{AF}$ is dominated by the change in the $K_{FS}$[15)]. These results indicate the importance of the estimating the $K_{FS}$. The underlying mechanism has not been completely clarified, but we have experimentally demonstrated an increase in the $T_B$ of $Cr_2O_3$ with increasing $a$, indicating increased $K_{AF}$ of $Cr_2O_3$.

### *4.3. Lattice strain effect on blocking temperature and Néel temperature*

Finally, we discuss the lattice strain effect on the $T_B$ and $T_N$. The $T_N$ data for these samples are included in table 2. As previously reported by us[26)], the $T_N$ of an α-$Fe_2O_3$-buffered sample is about 25 K less than that of a Pt-buffered sample, and the $T_N$ decreases with increasing $a$. There is a trade-off between the lattice strain effects on the $T_B$ and $T_N$: the $T_B$ increases with increasing $a$, while the $T_N$ decreases. This dependence appears to be unfavorable. However, the lattice strain effect is stronger for the $T_B$ than the $T_N$. Compared to the Pt-buffered sample, a 100 K higher $T_B$ was obtained for the α-$Fe_2O_3$-buffered sample, while the $T_N$ reduction was about 25 K. Thus, it is possible to increase the $T_B$ with only a small reduction in the $T_N$.

## 5. Conclusions

In this study, we discovered a high $T_B \approx 260$ K for a 20-nm-thick $Cr_2O_3$ thin film using a Ru spacer layer and an α-$Fe_2O_3$ buffer layer. The Ru spacer enabled reproducible control of the magnitude of the $H_{ex}$ ($J_K$), and by reducing the $H_{ex}$ ($J_K$) we enhanced the $T_B$. By changing the buffer layer material from Pt to α-$Fe_2O_3$, a higher $T_B$ was attained. The enhancement of the $T_B$ may be due to the lattice strain induced $K_{AF}$ change, which we estimate four times higher $K_{AF}$ for α-$Fe_2O_3$-buffered $Cr_2O_3$ film compared for Pt-buffered $Cr_2O_3$ film. We also clarified the trade-off between the $T_N$ and $T_B$ with respect to the lattice strain of $Cr_2O_3$, and demonstrated that the $T_B$ is more sensitive to lattice strain than the $T_N$. Such a control of the $T_B$ is the first step towards utilizing the ME effect in $Cr_2O_3$ thin films. Combined with further improvement of material properties, these techniques for controlling the $T_B$ open up doors for the device application.

**Acknowledgements**

This research was partly supported by the Advanced Low Carbon Technology Research and Development Program (ALCA) of the Japan Science and Technology Agency (JST), the Impulsing Paradigm Change through Disruptive Technologies Program (ImPACT) of the Council for Science, Technology and Innovation (Cabinet Office, Government of Japan), the Murata Science foundation, and a Grant-in-Aid from the Japan Society for the Promotion of Science (JSPS) Fellows.

Figures and Tables.

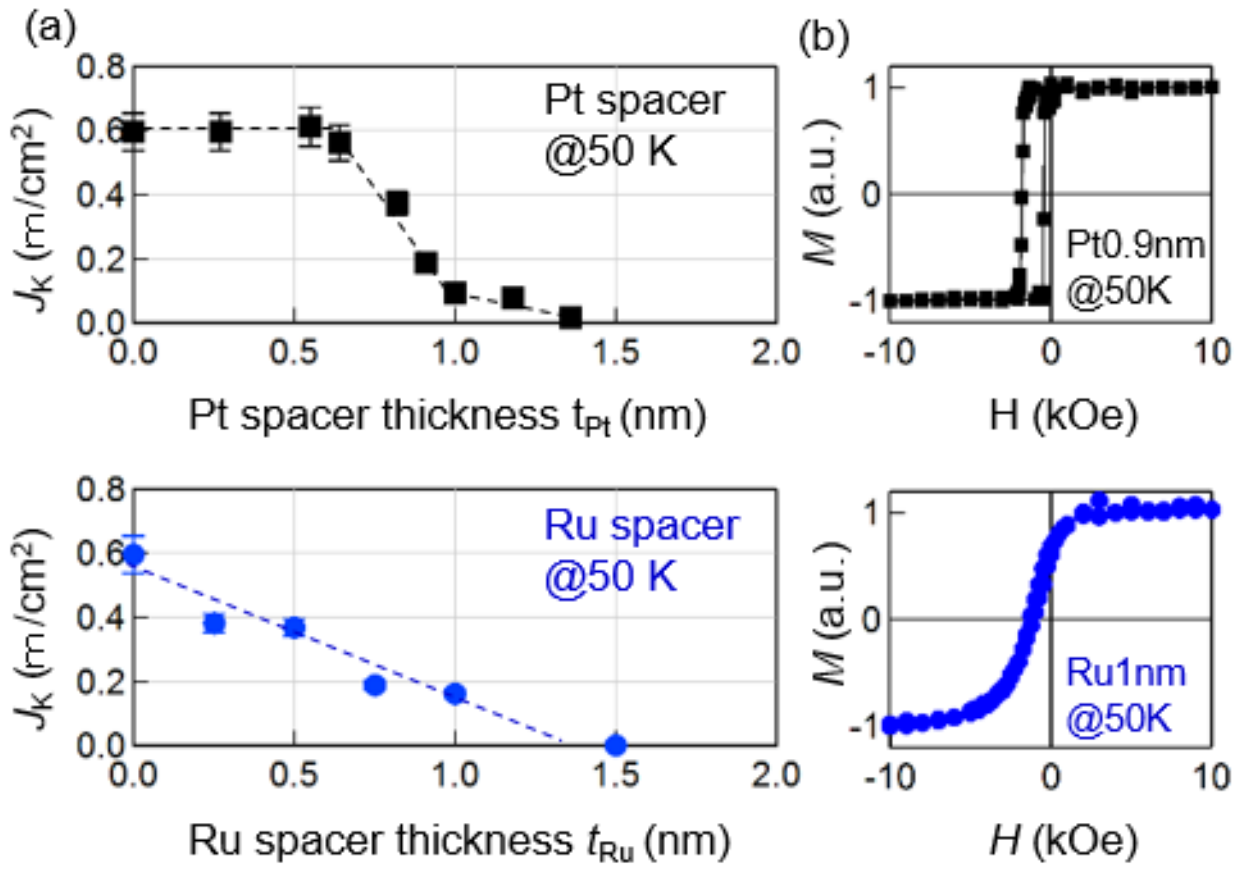

**Figure 1.** (a) Relationship between $J_K$ and spacer layer thickness for Ru and Pt spacer samples[22] at 50 K. The dashed lines are just guide for eyes. (b) *M*–*H* curves for Ru and Pt spacer samples at 50 K.

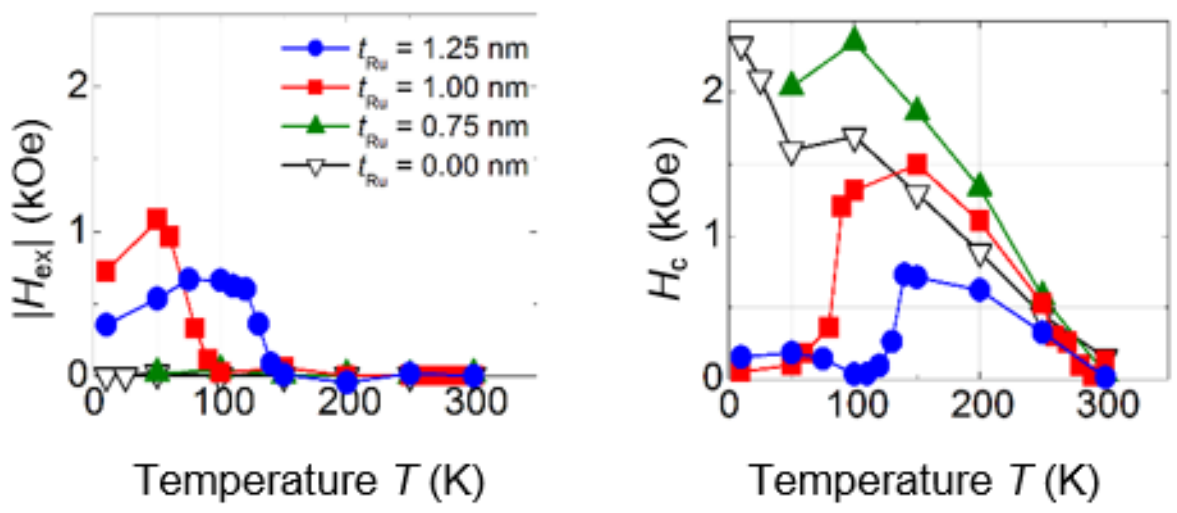

**Figure 2.** Temperature dependence of $|H_{ex}|$ (left panel) and $H_c$ (right panel) for c-$Al_2O_3$ substrate/Pt 25/$Cr_2O_3$ 20/Ru $t_{Ru}$/Co 1/Pt 5 (nm).

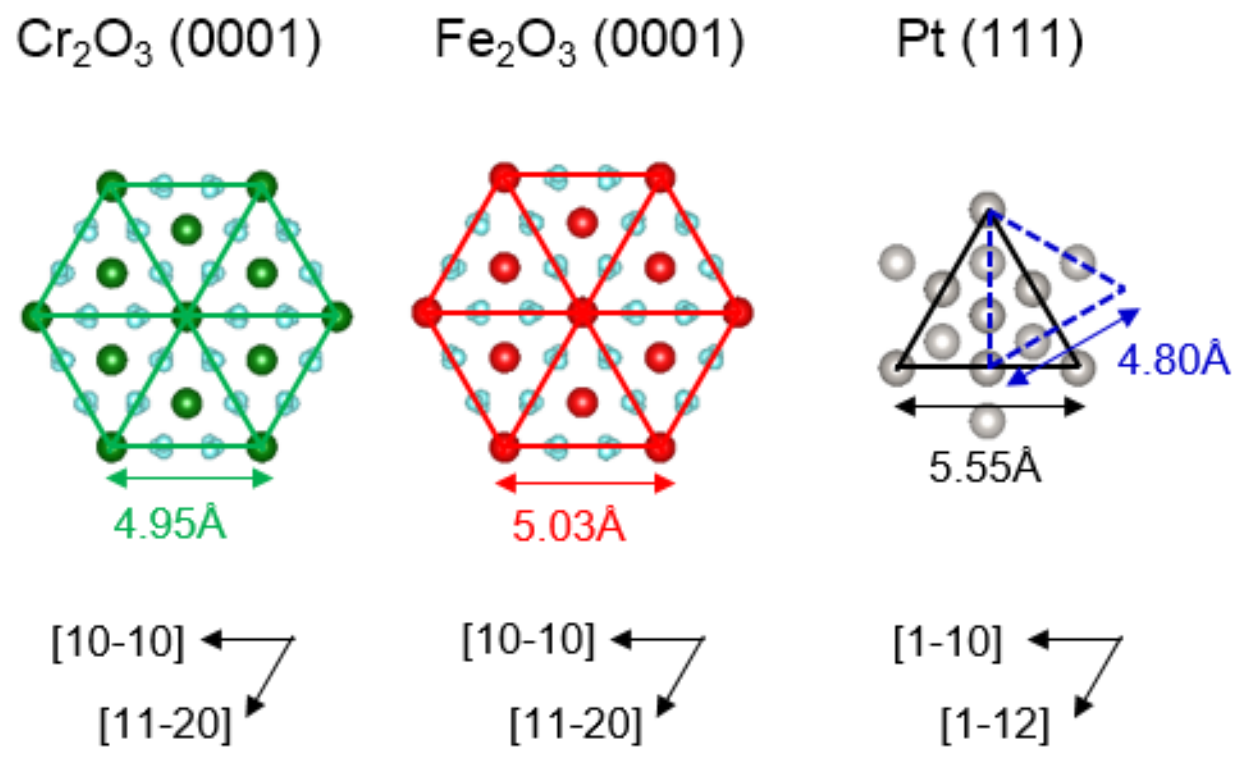

**Figure 3.** Schematic diagrams of the (0001) plane of $Cr_2O_3$, (0001) plane of α-$Fe_2O_3$, and (111) plane of Pt. The black solid line and blue broken line in Pt (111) plane indicate triangular lattice along [$1\bar{1}0$] and [$11\bar{2}$] direction, respectively.

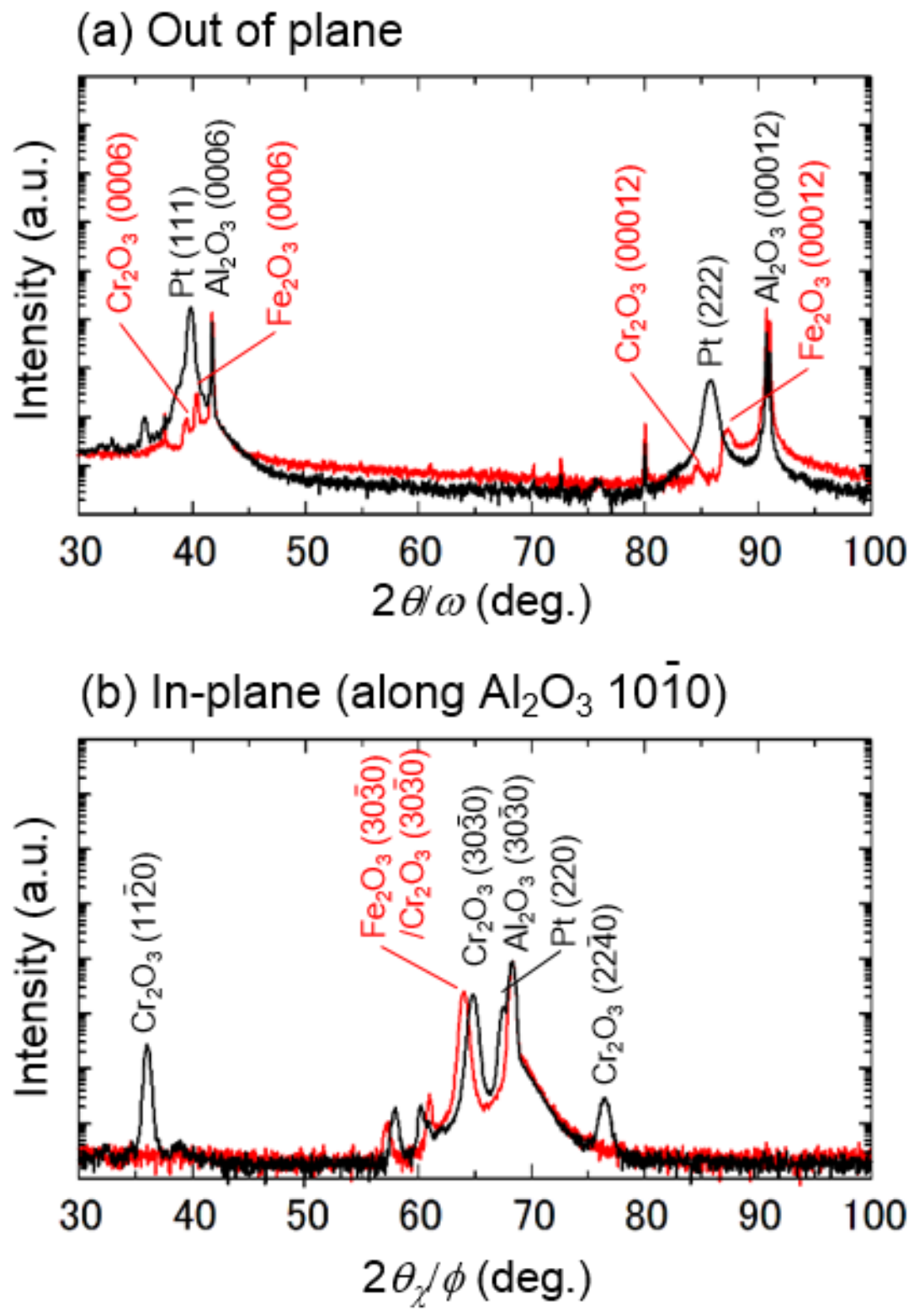

**Figure 4.** XRD patterns of (a) $2\theta/\omega$ (out of plane) and (b) $2\theta_\chi/\phi$ (in-plane) scans of α-$Fe_2O_3$- (red line) and Pt-buffered (black line) samples. The $2\theta_\chi/\phi$ scan was carried out for the $Al_2O_3$ $(10\bar{1}0)$ plane.

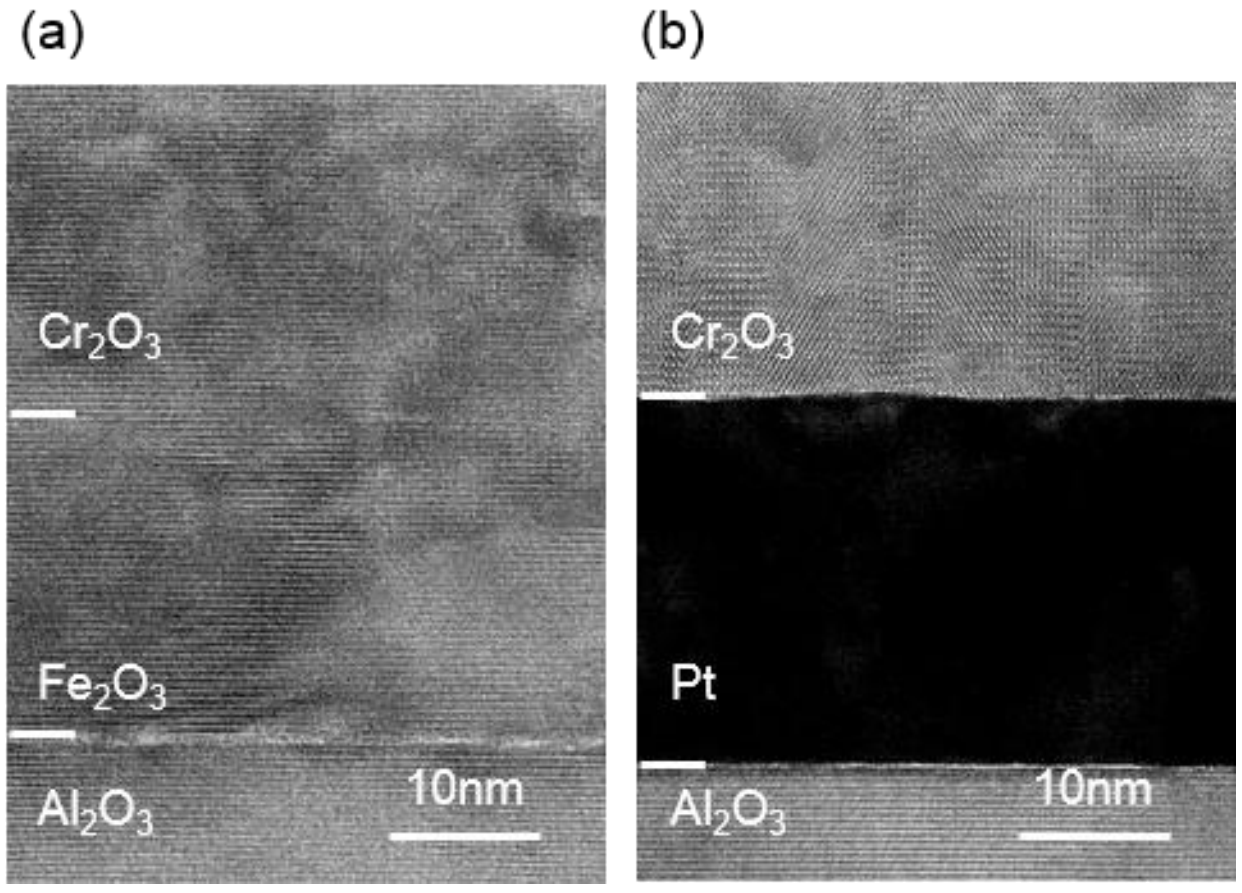

**Figure 5.** Cross-sectional TEM images of (a) α-$Fe_2O_3$- and (b) Pt-buffered samples.

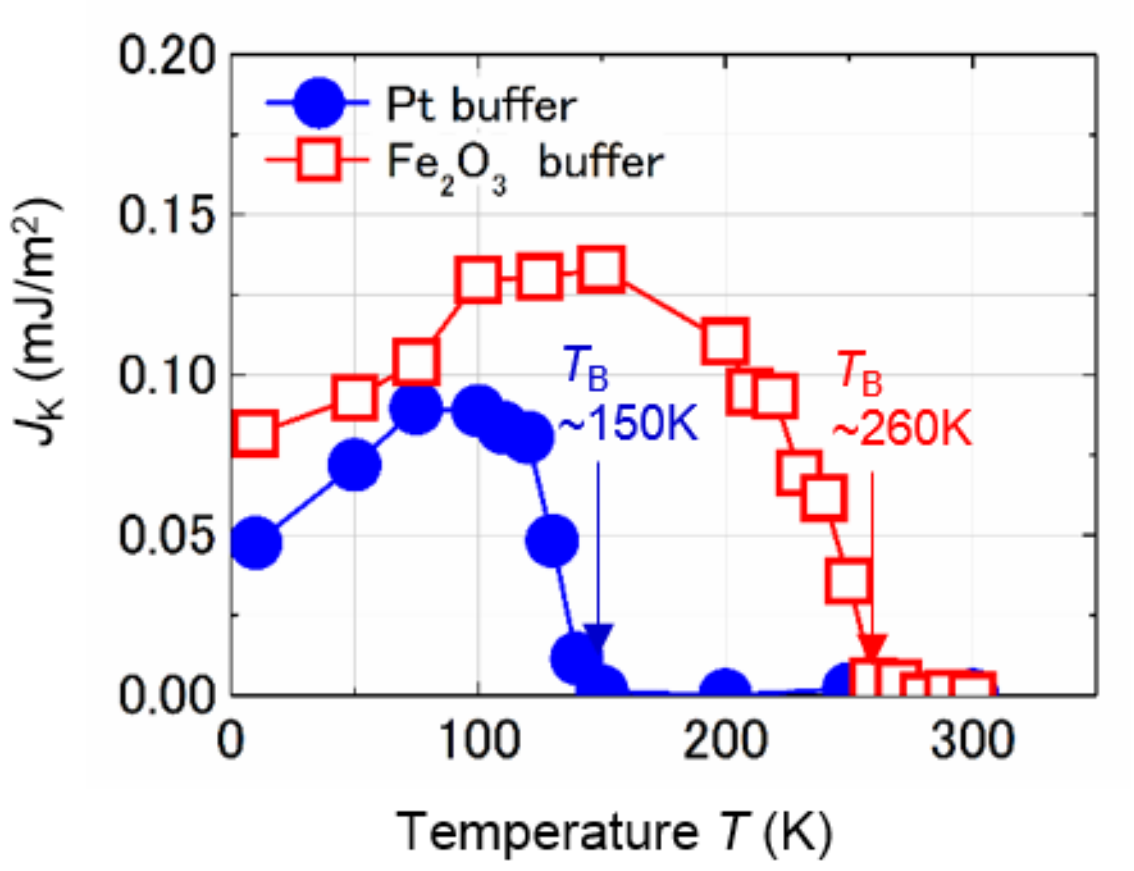

**Figure 6.** Temperature dependence of $J_K$ for Pt- (blue solid circles) and $\alpha$-$Fe_2O_3$-buffered (red open squares) 20-nm-thick $Cr_2O_3$ samples (c-$Al_2O_3$ substrate/Pt 25 or $\alpha$-$Fe_2O_3$ 20/$Cr_2O_3$ 20/Ru 1.25/Co 1/Pt 5 (nm)).

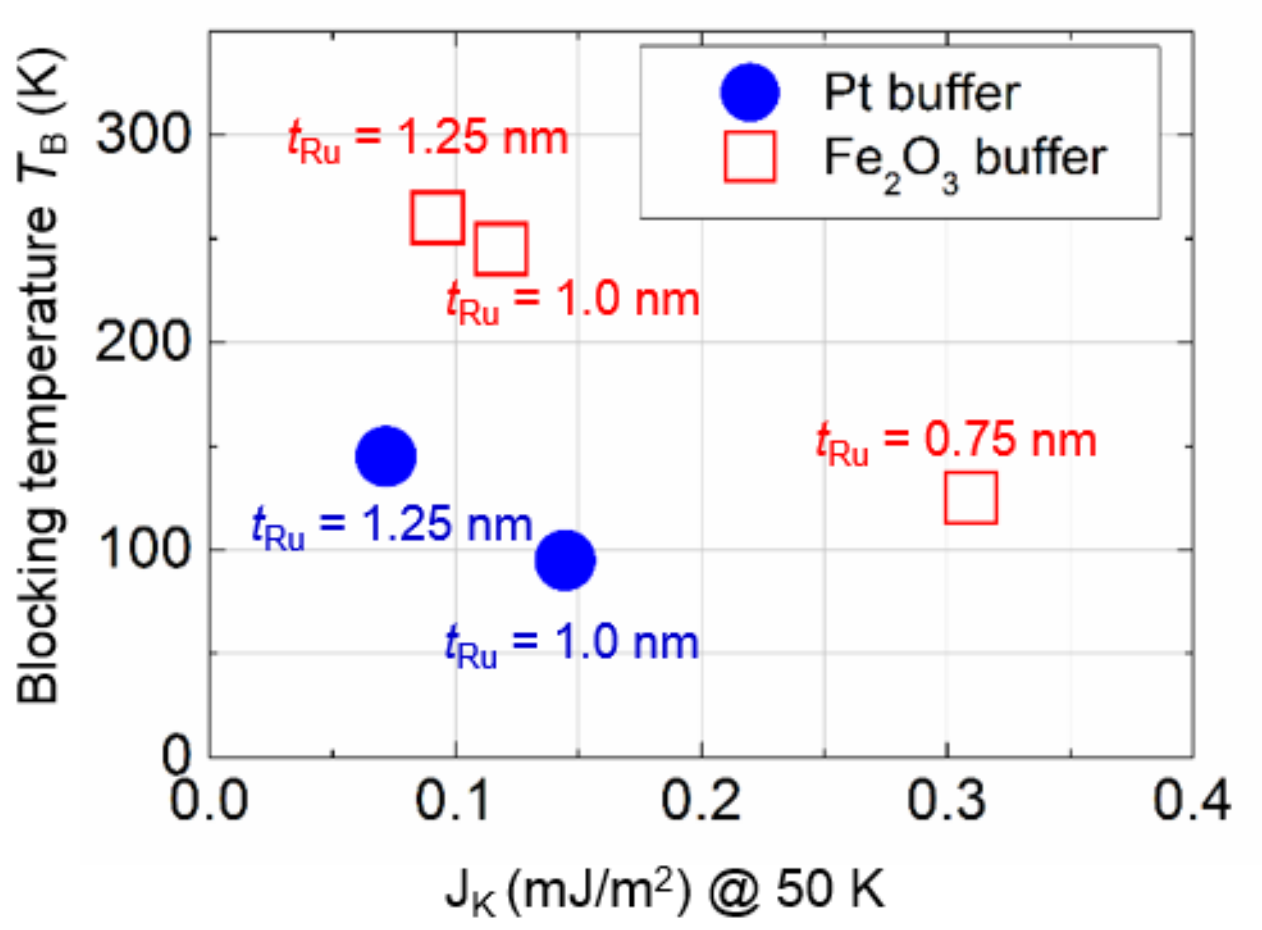

**Figure 7.** Relationship between $T_B$ and $J_K$ for Pt- and $\alpha$-$Fe_2O_3$-buffered 20-nm-thick $Cr_2O_3$ samples (c-$Al_2O_3$ substrate/Pt 25 or $\alpha$-$Fe_2O_3$ 20/$Cr_2O_3$ 20/Ru $t_{Ru}$/Co 1/Pt 5 (nm)) at 50 K.

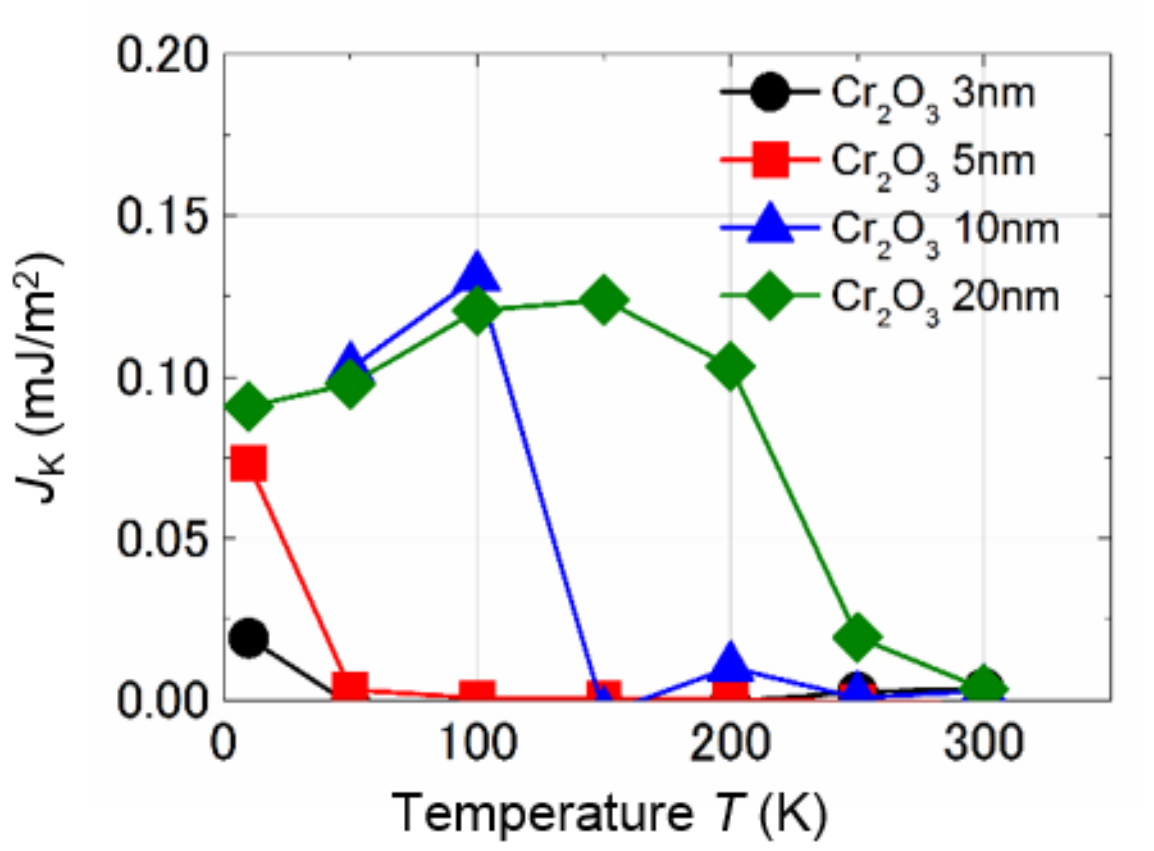

**Figure 8.** Temperature dependence of $J_K$ for $Al_2O_3$/$\alpha$-$Fe_2O_3$ 20/$Cr_2O_3$ $t_{Cr2O3}$/Ru 1.25/Co 1/Pt 5 (nm) with various $Cr_2O_3$ thicknesses ($3 \leq t_{Cr2O3} \leq 20$).

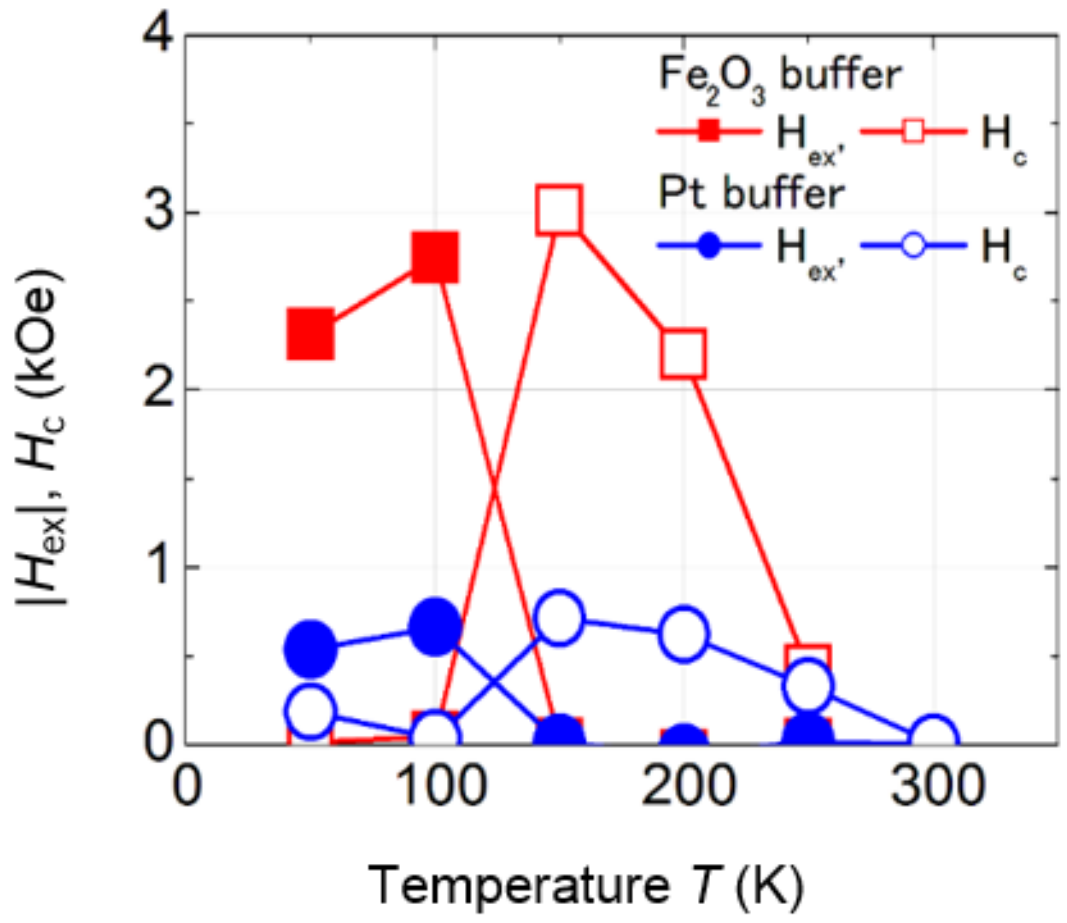

**Figure 9.** Temperature dependence of $H_{ex}$ and $H_c$ for $Al_2O_3$/α-$Fe_2O_3$ 20/$Cr_2O_3$ 20/Ru 0.75/Co 1/Pt 5 ($Fe_2O_3$ buffer) and $Al_2O_3$/Pt 25/$Cr_2O_3$ 20/Ru 1.25/Co 1/Pt 5 (Pt buffer).

**Table 1.** Lattice mismatch between $Cr_2O_3$ bulk and buffer layer materials, and experimental *a* value of $Cr_2O_3$ layer.

| Buffer | *a* of buffer (difference from *a* of bulk $Cr_2O_3$) | | Orientation | *a* of $Cr_2O_3$ layer from XRD (difference from *a* of bulk $Cr_2O_3$) |
|---|---|---|---|---|
| ($Cr_2O_3$) | 4.95 Å | (-) | | |
| Pt | 4.80 Å | (−3.1%) | Pt [110]//$Cr_2O_3$ $[11\bar{2}0]$ | 4.98 Å (+0.44%) |
| α-$Fe_2O_3$ | 5.04 Å | (+1.8%) | α-$Fe_2O_3$ $[10\bar{1}0]$//$Cr_2O_3$ $[10\bar{1}0]$ | 5.04 Å (+1.69%) |

**Table 2.** Lattice parameter *a* of the $Cr_2O_3$ layer, and $T_B$ and $T_N$[26] of Pt- and α-$Fe_2O_3$-buffered 20-nm-thick $Cr_2O_3$ samples (c-$Al_2O_3$ substrate/Pt 25 or α-$Fe_2O_3$ 20/$Cr_2O_3$ 20/Ru 1.25/Co 1/Pt 5 (nm)).

| Buffer | *a* of $Cr_2O_3$ layer from XRD (difference from *a* of bulk $Cr_2O_3$) | | $T_B$ | $T_N$ [26] |
|---|---|---|---|---|
| Pt | 4.98 Å | (+0.44%) | 150 K | 294 K |
| α-$Fe_2O_3$ | 5.04 Å | (+1.69%) | 260 K | 269 K |